\documentclass[journal=jacsat,manuscript=article]{achemso}

\usepackage[version=3]{mhchem} 

\usepackage{xcolor}
\usepackage{gensymb}

\usepackage{tablefootnote}
\usepackage[normalem]{ulem}

\title{Crystal Growth Characterization of \ce{WSe2} Thin Film Using Machine Learning}
\author{Isaiah A. Moses}
\affiliation[MRI]{Materials Research Institute, The Pennsylvania State University, University Park, PA 16802}
\author{Chengyin Wu}
\affiliation[MTSE]{Department of Materials Science and Engineering, The Pennsylvania State University, University Park, PA 16802}
\author{Wesley F. Reinhart}
\email{reinhart@psu.edu}
\affiliation[MTSE]{Department of Materials Science and Engineering, The Pennsylvania State University, University Park, PA 16802}
\alsoaffiliation[ICDS]{Institute for Computational and Data Sciences, The Pennsylvania State University,
University Park, PA 16802}

\begin{document}

\maketitle
\begin{abstract}
Materials characterization remains a labor-intensive process, with a large amount of expert time required to post-process and analyze micrographs.
As a result, machine learning has become an essential tool in materials science, including for materials characterization.
In this study,  we perform an in-depth analysis of the prediction of crystal coverage in \ce{WSe2} thin film atomic force microscopy (AFM) height maps with supervised regression and segmentation models.
Regression models were trained from scratch and through transfer learning from a ResNet pretrained on ImageNet and MicroNet to predict monolayer crystal coverage.
Models trained from scratch outperformed those using features extracted from pretrained models, but fine-tuning yielded the best performance, with an impressive 0.99 $R^2$ value on a diverse set of held-out test micrographs.
Notably, features extracted from MicroNet showed significantly better performance than those from ImageNet, but fine-tuning on ImageNet demonstrated the reverse.
As the problem is natively a segmentation task, the segmentation models excelled in determining crystal coverage on image patches.
However, when applied to full images rather than patches, the performance of segmentation models degraded considerably, while the regressors did not, suggesting that regression models may be more robust to scale and dimension changes compared to segmentation models.
Our results demonstrate the efficacy of computer vision models for automating sample characterization in 2D materials while providing important practical considerations for their use in the development of chalcogenide thin films.
\end{abstract}



\section{Introduction}
Great advances are being made in the synthesis of two-dimensional materials (2D)\cite{gupta2015recent, mas20112d,lv2015transition}, since the successful isolation of graphene in 2004\cite{novoselov2004electric}.
The transition metal dichalcogenides (TMD) is a major class of 2D materials that have gained much attention due to their interesting properties and potential for applications in areas including electric and optoelectronic, energy, and sensing\cite{lv2015transition, choi2017recent}.
Several synthesis methods, including mechanical exfoliation\cite{lv2015transition}, powder vaporization\cite{huang2014large, lin2014direct}, pulsed laser deposition\cite{grigoriev2012experimental}, chemical vapor deposition (CVD), and metal-organic chemical vapor deposition (MOCVD)\cite{1eichfeld2015highly, 5zhang2016influence, kang2015high, 7kim2017suppressing} are being deployed in a bid to improve both the quality and scalability of the grown  TMDs.
Associated with the materials synthesis is the need for an efficient characterization technique to determine the various features of the samples, ranging from the basic crystal qualities to the determination of the properties and potential applications of the materials\cite{lin2018realizing, lin20162d}.

Atomic force microscopy (AFM) is a scanning probe microscopy that is widely applied in 2D materials characterization due to its versatile capability in the electrical, mechanical, chemical, thermal, electrochemical, and topological characterization of samples\cite{rugar1990atomic, giessibl2003advances,zhang2018atomic}.
The topological mode of the AFM is crucial in determining the quality and properties of a sample as it is used to produce an AFM image from which several characteristics, including crystal coverage, domain size, shape and thickness, and nucleation density can be determined\cite{8cohen2020growth, 9cun2019wafer, 5zhang2016influence, 2li2021epitaxial, 3xiang2020monolayer}.
Given the fundamental role the information from the AFM image analysis plays in determining the grown sample's quality, even before further characterization to determine their properties and potential applications, the fidelity and efficiency of the analysis are of major priority in the workflow to accelerate the 2D materials qualitative and quantitative synthesis and exploration.

The application of machine learning in image analysis, particularly in segmentation, is an important and actively researched area in materials science and related fields.\cite{stuckner2022microstructure, seg1akers2021rapid, 10borodinov2020machine, Unet++_zhou2019unet++, seg2azimi2018advanced} This interest stems from its potential to automate processes, reduce human intervention, and swiftly handle large volumes of images. Numerous software tools, such as ImageJ,\cite{abramoff2004image} Gwyddion,\cite{nevcas2012gwyddion} WSxM,\cite{horcas2007wsxm} NanoScope Analysis,\cite{abramoff2004image} Mountain,\cite{Mountains} and MIPAR,\cite{sosa2014development} have been developed to address image correction, processing, and analysis needs. Some of these tools support automation and batch analysis, enabling the processing of many images. Recently, deep learning strategies have also become a mainstay of this field, and some software, like MIPAR, now support deep learning workflows natively. This raises questions about the performance and reliability of machine learning schemes for materials characterization. Specifically, there is a need to explore the behavior of regression models compared to segmentation models for characterizing micrographs, such as determining the thin film crystal growth on a substrate, quantified by crystal coverage. Furthermore, investigating how pretraining domains and different modes of transfer learning impact the capabilities and reliability of models at inference time is crucial. Analyzing these aspects will contribute to a better understanding of the overall potential of different learning schemes and, more specifically, their suitability for high-throughput characterization. This is essential for accelerating the exploration of TMDs.

Several studies have been reported on the deployment of ML models to the AFM image analysis.
Among them are the segmentation of the molecular resolved AFM images\cite{10borodinov2020machine}, classification of quasi-planar molecules that spans relevant structural and compositional moieties in organic chemistry based on AFM images\cite{p12carracedo2021deep}, identification of self-organized nanostructures\cite{11gordon2020automated}, extraction of molecule graphs of samples from AFM images\cite{p13oinonen2022molecule}, atomic structure recovery from AFM images\cite{12alldritt2020automated}, and quantitative analysis of \ce{MoS2} thin film micrographs.\cite{moses2023quantitative}
Crucial to the determination of the quality of the materials synthesis is the domain size and thickness, and surface coverage\cite{6tang2023migration, 7kim2017suppressing, 8cohen2020growth, 9cun2019wafer, bachu2023role, chen2023large}, isolation of the grown crystal from the substrate on which it is grown. 

The crystal coverage is a basic metric that indicates the extent to which the thin film has grown on the substrate.
A rapid and automated determination of the crystal coverage can enhance materials synthesis as the growth parameters can be optimized based on this figure of merit.
In our present study, convolutional regression models are developed to be deployed in determining the crystal coverage of 2D \ce{WSe2} grown using MOCVD\cite{1eichfeld2015highly}.
Additionally, robust semantic segmentation models\cite{seg3holm2020overview, seg4zhao2023new, seg5baskaran2020adaptive, seg6kim2020unsupervised, seg7gupta2020modelling} which give a pixel-wise classification of the grown samples AFM images, as either belonging to the substrate or the crystals, are trained.
Our models exhibit excellent results with $R^2$ exceeding 0.99 in the quantification of the crystal coverage in held-out test samples.

Furthermore, we have systematically evaluated the efficacy of different transfer learning schemes, namely feature extraction and fine-tuning.
We also include the effects of different pretraining domains, specifically materials micrographs compared to miscellaneous everyday objects.
Our results have some important and counter-intuitive implications for the practical implementation of these computer vision models in materials characterization workflows.

\section{Method}
\subsection{Dataset}


\begin{figure}
     \centering
     \includegraphics[width=\textwidth]{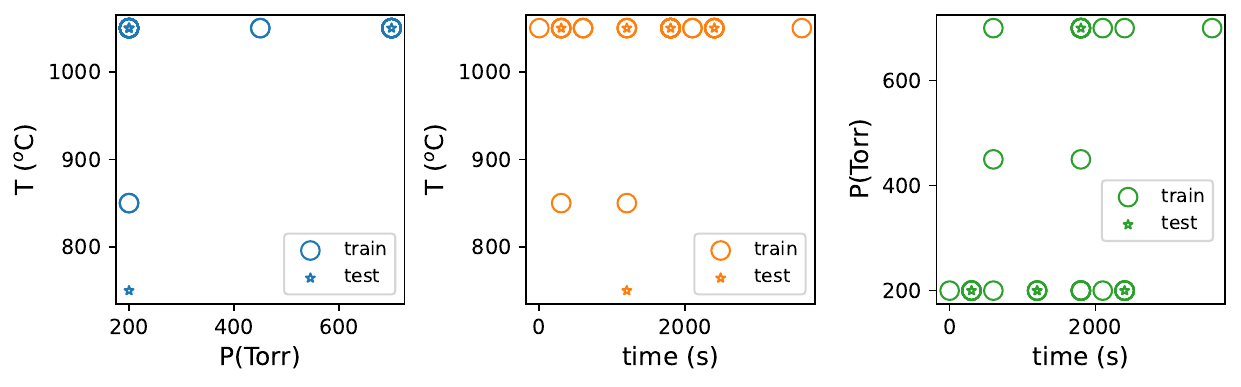}
     \caption{\ce{WSe2} samples used in the study showing the growth parameters space.
     $T$ is the growth chamber inner temperature, $P$ is the pressure, and time is the growth time. Multiple micrographs are obtained for each sample, so there are fewer unique conditions than images in our study.
     Bolder circles indicate more samples at the same point.
     Some samples in the test set occupy unique points in the parameter space, such as the samples at the lowest $T$.} 
     \label{fig:data}
\end{figure}

\begin{figure}
     \centering
     \includegraphics{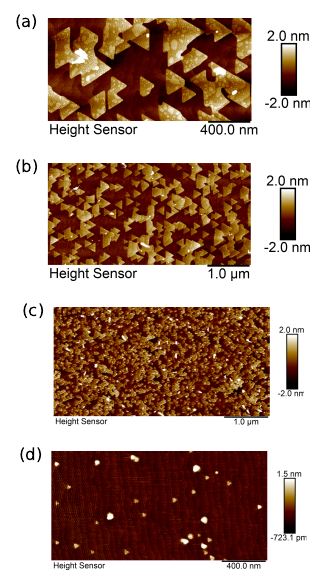}
     \caption{Sample AFM images of \ce{WSe2} thin film in our dataset.
     Data ingested in our workflow have already been preprocessed by other software and include dimensional scale bar, color scale, and text annotations.} 
     \label{fig:AFM}
\end{figure}

The \ce{WSe2} AFM data used in this research were grown by Eichfeld et al.\cite{1eichfeld2015highly} and stored in the Lifetime Sample Tracking (LiST), a database hosted by the 2D Crystal Consortium (2DCC)\cite{WSe2data}, while the processed data are available to download from Ref.\cite{moses_2024_10784189}.
The 52 \ce{WSe2} thin film samples were synthesized using the metal-organic chemical vapor deposition (MOCVD) technique.
The samples were grown at various conditions, including the growth time, chamber inner temperature, and pressure (Fig.~\ref{fig:data}), resulting in significant variations in the morphological features of the AFM micrographs obtained.
Additionally, different imaging conditions were employed for the samples, with characterization obtained at the centers and edges of the wafer and different resolutions.
This resulted in a total of 221 micrographs from the 52 grown samples.

Eichfeld et al.,\cite{1eichfeld2015highly} which grew the samples, postprocessed the micrographs using NanoScope Analysis\cite{abramoff2004image} that performed flattening (which we could have done automatically as part of our workflow), inserted a color bar, and annotated the images with text labels and a scale bar. We therefore retrieved flattened images which were stored as TIF files such as those shown in Fig.~\ref{fig:AFM}.
One important consequence of using the flattened images is that our models were trained not on height maps, but on height-normalized images.
That is, the relationship between pixel intensity and the original height measurement was different within each image.
The same was true of the length scale, where pixels represented different sample areas within each image.
We believe this better represents the practical use case for these models compared to carefully controlled height and length scales. For the deployment of our models on raw (unflatten) images, the flattening step can be implemented as part of an automated workflow (e.g., pySPM\cite{olivier_scholder}).

\begin{figure}[h!]
    \centering
    \includegraphics[width=\textwidth]{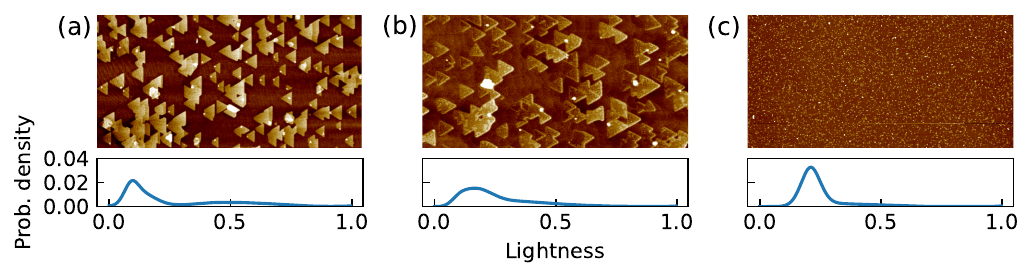}
    \caption{Sample images with their lightness histograms demonstrating how segmentation could be performed based on a bimodal lightness distribution (one for foreground, one for background).
    This assumption is often violated due to imperfect flattening, texture, or artifacts.}
    \label{fig:histograms}
\end{figure}

The figure of merit for these thin film samples is the monolayer coverage, which can be computed from an AFM height map according to the fraction of pixels in the foreground compared to the overall image.
This essentially reduces the problem to a segmentation task, which has many possible solutions.
One simple method to perform binary segmentation (i.e., foreground/background separation) is to define a lightness threshold (corresponding to a height threshold) based on the assumption of a mixture of approximately Normal distributions for each height range of interest (such as background and foreground). This approach however has limitations as the Normal distribution assumption is often violated due to imperfect flattening, texture, or artifacts.
With a script that was applied to all the images, each image was cropped to only the AFM micrograph portion (no padding, annotations, color bar, scale bar, etc.), a lightness histogram was prepared, and a threshold value was selected based on an assumed bimodal distribution, as shown in Fig.~\ref{fig:histograms}.
Choosing this threshold produces a binary mask for each image; these thresholds were chosen and masks evaluated manually for each micrograph.
This labeling procedure resulted in 221 image-mask pairs, from which the monolayer coverage was computed by counting the number of pixels above the lightness threshold (i.e., masked). 

\subsubsection{Augmentation}

\begin{figure}[h!]
        \centering
        \includegraphics{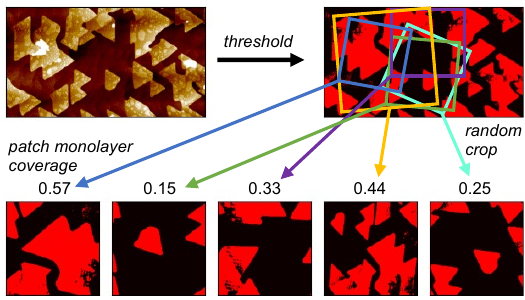}
        \caption{Data augmentation and image patching schematic scheme.
        The original AFM image (top left) is thresholded to produce a mask (top right).
        Random image patches are jointly taken from both the image and mask to yield new (image, mask, coverage) sets where all patches are of size $224 \times 224$ but represent different portions of the original image.
        }
        \label{fig:ImagePatch}
    \end{figure}

A dataset consisting of only 221 images might be insufficient to effectively train a robust ML model.
Therefore, in this study, we utilized image patching, a common data augmentation technique to generate additional data points with greater variance in image characteristics, thus creating a more diverse dataset for deep learning model training.
We utilized the random transforms implemented in torch-vision from the Pytorch library\cite{PYTORCHpaszke2019pytorch} to generate the image patches, with a final patch height and width of $224 \times 224$ for regression models and $512 \times 512$ for the segmentation models.
Each patch had an equal and independent chance of being flipped vertically, horizontally, $0-360^\circ$ rotation, $0.5 - 2.0 \times$ rescale, and random crop within the rescaled image.
An example of this procedure is shown in Fig.~\ref{fig:ImagePatch}.
Because this random transformation could result in out-of-bounds pixels, we rejected any patch that did not fall entirely within the original image.
We repeated this sampling until 10 valid patches were obtained for each image.

\subsection{Regression Models}

\begin{figure}
        \centering
        \includegraphics{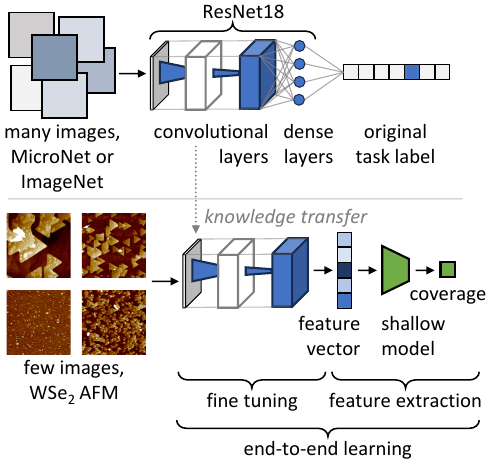}
        \caption{Schematic of different transfer learning paradigms.
        Feature extraction is a scheme that only modifies the trainable weights in the fully connected layer (or other shallow models) while leaving the pretrained weights in the convolutional layers unchanged.
        In fine tuning, all the trainable weights from the pretrained model are adjusted to improve the model's fit to the new task.
        In end-to-end learning, the entire model is trained from scratch, without any knowledge transfer.
        }
        \label{fig:TL}
    \end{figure}

We consider two variants of the ML task: regression (predicting the coverage label directly from the image) and segmentation (predicting the binary mask and then computing the coverage from the mask).
Within the regression task, we further consider three training paradigms: training from scratch using end-to-end learning (i.e., with randomly initialized weights), transfer learning by fine-tuning (i.e., initializing the model with pretrained weights), and transfer learning by feature extraction (i.e., training a shallow model to predict target label with pretrained convolutional filters) (see Fig. \ref{fig:TL}). 

For all the regression models, the Adam optimizer, ReLU activation function, and mean squared error (MSE) loss functions were used.
10\% of the data samples, grown under different growth parameters than the rest of the data and/or obtained under different imaging conditions, were held out to determine how well the models generalize to out-of-distribution data (Table~\ref{fig:data}).
Additionally, about 80\% and 10\% were used for the training and validation, respectively.

We started by training a small Convolutional Neural Network from scratch (CNNsc).
The architecture of the CNNsc network was optimized using Bayesian hyperparameter tuning implemented in the \texttt{ax-platform} package\cite{bakshyopen} which leverages a Gaussian-process-based Bayesian optimization\cite{snoek2012practical}..
After each of the convolutional layers, a max pooling and ReLU activation function were applied to downsize the feature maps and extract the most important features, and introduce non-linearity, respectively.
This network was deliberately simplified compared to the pretrained models to evaluate whether fewer trainable weights would be more robust in extrapolating to the test domain.

We also explored the application of pretrained models, specifically ResNet18 architecture pretrained on ImageNet\cite{5206848} and MicroNet\cite{stuckner2022microstructure} datasets, to predict the coverage of \ce{WSe2} thin films.
We chose ResNet18 as it is among the shallowest standard computer vision architectures available today, which we felt was important given our low data volume.
The features were extracted from the average pool layer of the pretrained models, given 512 features.
Multilayer perception (MLP) models were then built to learn the crystal coverage from the image features obtained from the ResNet18 pretrained on the ImageNet and MicroNet.
The MLP models are hereafter referred to as MLP-I and MLP-M, respectively.
MLP model hyperparameters were tuned using \texttt{ax-platform} as in the case of the CNNsc.

For completeness, we also employed the fine-tuning paradigm of transfer learning.
This allowed us to assess the performance of these pretrained models in our specific context and evaluate their potential for accurate thin film coverage prediction.
The pretrained models' classifiers were replaced with 2 FC (fully connected) layers of 512 and 100 neurons and an output layer.
Between the 2 FC layers is a ReLU activation function to introduce non-linearity and a dropout of 0.25  to minimize over-fitting.
The sigmoid activation function was additionally placed before the output layer to ensure only values between 0.0 and 1.0 (range of coverage values) are predicted.
The models were then tuned with our data to learn the crystal coverage.
The fine-tuning was carried out for the ResNet18 pretrained on the ImageNet (CNN-I) and another on the MicroNet (CNN-M).

\subsection{Segmentation Models}

Separately from the regression task, we attempt to solve the problem using segmentation models to work natively with the binary mask.
Similar to the regression models, encoders pretrained on MicroNet by Stuckner et al. \cite{stuckner2022microstructure} were used.
In their report, they found ResNeXT,\cite{ResNeXT_xie2017aggregated} SE,\cite{SE_hu2018squeeze} Inception,\cite{Inception_szegedy2017inception} and EfficientNet\cite{EfficientNet_tan2019efficientnet} encoder architectures to give better performances.
Additionally, Unet\cite{Unet_ronneberger2015u} and Unet++\cite{Unet++_zhou2019unet++} decoders were found to outperform others.
Specifically, \texttt{SE\_ResNeXt-50\_32x4d} and \texttt{SE\_ResNeXt-101\_32x4d} encoders pretrained on MicroNet coupled with Unet++ decoders gave, on the average, the best intersection over union (IoU) accuracy for models trained on the full sets of 2 different SEM images (nickel-based superalloys and environmental barrier coatings).
We therefore used \texttt{SE\_ResNeXt-50\_32x4d} and \texttt{SE\_ResNeXt-101\_32x4d} encoders pretrained on MicroNet coupled with Unet++ decoders in our study.
These segmentation models are termed SEG50 and SEG101, respectively. 

For us to compare the performance of the segmentation and regression models from the same pretrained architectures, we have additionally trained segmentation models based on the ResNet18 pretrained encoder and using the Unet++ decoder.
Both encoders pretrained on the ImageNet and MicroNet were used, and termed SEG18-I and SEG18-M, respectively.
The Adam optimizer, 1e-4 learning rate, and a batch size of 6 were used in the training.
We utilized an early stopping after 30 epochs of training without further improvement on the IoU accuracy of the validation set, while the loss function was a weighted sum of balanced cross entropy (BCE) and dice loss with a 70\% weighting towards BCE.

\section{Results and Discussion}

\subsection{Regression Models}

\subsubsection{Training from Scratch}

\begin{figure}
     \centering
     \includegraphics[width=\textwidth]{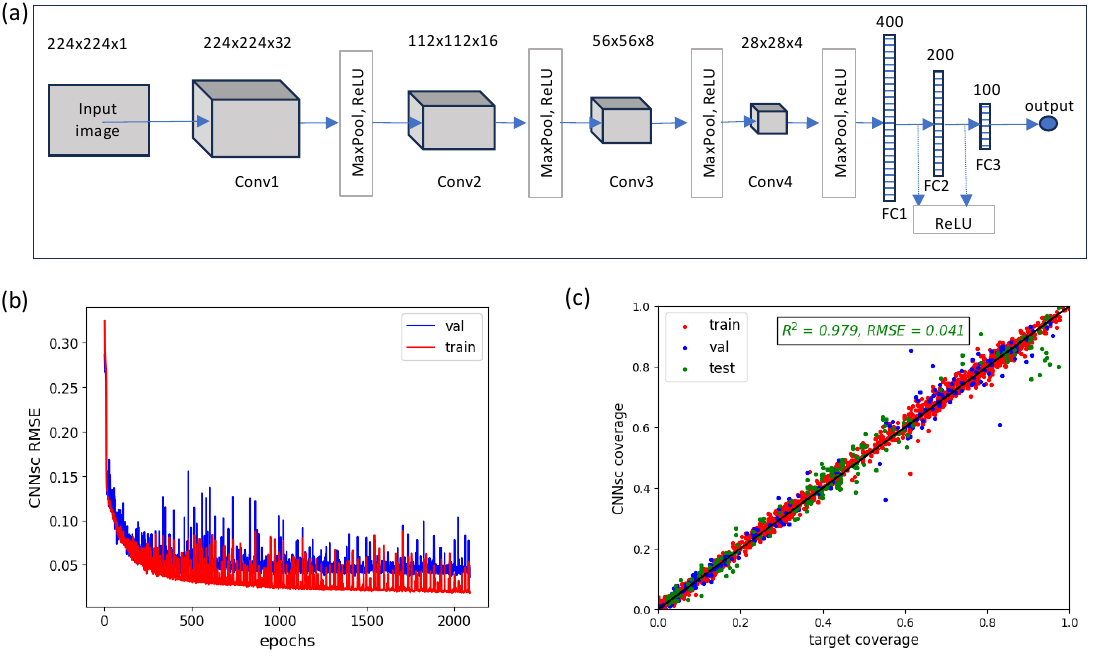}
         \caption{(a) is the CNN architecture built from scratch (CNNsc) showing the convolutional (Conv1, Conv2, Conv3, and Conv4), the pooling (MaxPool), and fully connected layers (FC1, FC2, FC3), as well as the feature maps and channel sizes for each of the convolution layer and the neurons connecting the FC layers.
         (b) is the root mean squared error value (RMSE) on the train and validation (val) data against the learning iteration (epochs).
         (c) is the parity plot of the predicted and target coverage. The $R^2$ and RMSE values in (c) are for the test set.}
         \label{fig:scratch}
\end{figure}

The architecture of the CNNsc network found by hyperparameter tuning consisted of four convolutional layers and three fully connected (FC) layers (Fig.~\ref{fig:scratch}).
The kernel size was 5 with a stride of 1 and zero padding.
This model was trained to minimize the MSE loss between the target and the predicted coverage.
A stochastic behavior is observed in the learning resulting in the fluctuation in losses with the training iterations both for the training and validation set (Fig.~\ref{fig:scratch}(b)).
The random initialization of the weights might have resulted in such behavior.
To obtain an optimally trained model, the model was set to stop once the minimal obtainable value of the training and validation loss was achieved.
This results in the model's performance with train, validation, and test set RMSE of 0.018, 0.039, and 0.041, respectively (Fig.~\ref{fig:scratch}(c) and Table~\ref{table:regression_results}).
These correspond to $R^2$ values of 0.997, 0.984, and 0.979 for train, validation, and test, respectively.
Only a few scattered points were observed in the validation and test parity plots, indicating a minimal over-fitting.

\subsubsection{Feature Extraction}

\begin{figure}
     \centering
     \includegraphics[width=0.9\textwidth]{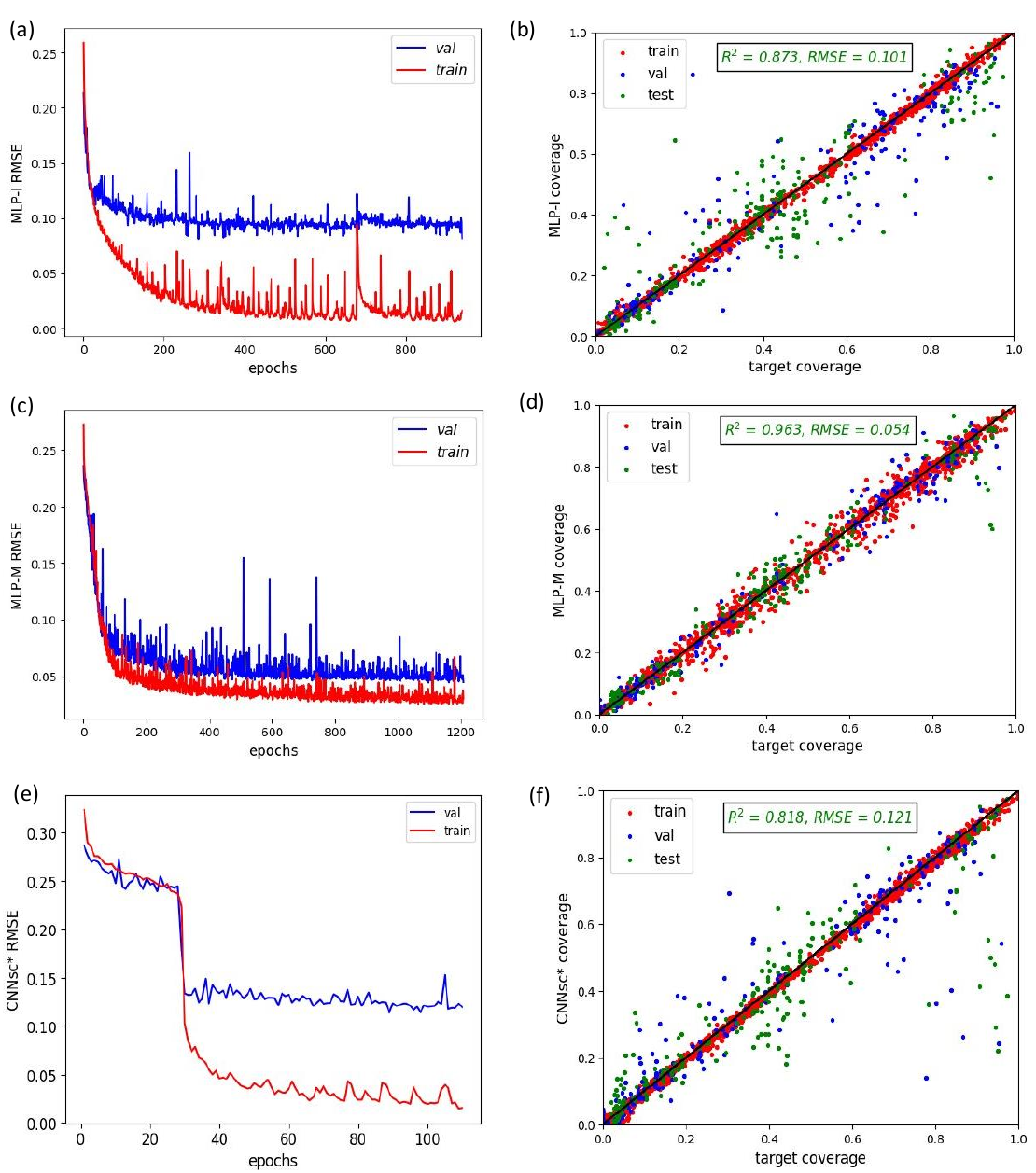}
         \caption{(a), (c), and (e) are the root mean squared error value (RMSE) on the train and validation (val) data against the learning iteration (epochs) for the multilayer perceptron model (MLP) trained with features extracted using the ResNet18 pretrained on the ImageNet data (MLP-I), MLP trained with features extracted using the ResNet18 pretrained on the MicroNet data (MLP-M), and for the CNN built from scratch without on-the-fly data augmentation (CNNsc*), respectively. (b), (d), and (f) are the parity plots of the predicted and target coverage corresponding to (a), (c), and (e), respectively. The $R^2$ and RMSE values in (b), (d), and (f) are for the test set.}
         \label{fig:extract}
\end{figure}

The MLP architectures were tuned (to minimize the validation loss) to yield 2 hidden layers with (120, and 84) neurons in the MLP-I and MLP-M.
The trained MLP-I exhibited an $R^2$ value of 0.873 on the test set (Fig.~\ref{fig:extract} and Table~\ref{table:regression_results}).
MLP-M performs better than the MLP-I, though still slightly worse than the CNNsc.
A better performance observed in the MLP-M than the MLP-I might be due to the proximity of the data for the pretraining and our data; MicroNet consists of grayscale micrographs while ImageNet is made up of the macroscale color images of natural objects.
The features extracted from the former may therefore be more relevant in learning our image features than those from the latter.

The superior performance of the CNNsc may be due to its smaller size or its on-the-fly data augmentations; random rotations and flips were applied to the data while training.
To verify if the data augmentations applied to the CNNsc made a significant difference to the model performance, we trained the same architecture of CNN with the same hyperparameters without the augmentations (CNNsc*).
The result shows that the augmentations indeed significantly enhance the performance of the CNNsc (Fig.~\ref{fig:extract} and Table~\ref{table:regression_results}).
Overfitting is observed to set in soon after the first few epochs of training on data without augmentation.
The model accurately predicts the coverage for the train set but a worse performance than both MLP-I and MLP-M is observed in the validation and test sets.

However, the on-the-fly augmentation cannot be readily applied in the feature extraction case as data are not seen by the model more than once.
The closest we can get to the on-the-fly augmentation is to obtain different features for the rotated and horizontal and vertically flipped images, then train the MLP model on all of these at once.
We also tried average pooling on these variants as input to the model rather than trying to learn a many-to-one mapping.
Both of these approaches gave worse performance compared to the vanilla MLP models, with the augmentation giving the $R^2$ values of 0.86 for the MLP-I and 0.93 for the MLP-M, while the pooling strategy was worse.
These results underscore a fundamental difference in the static augmentation of the data for the MLP models and the on-the-fly augmentation for the CNN models.

\subsubsection{Fine-Tuning}

\begin{figure}
     \centering
     \includegraphics[width=\textwidth]{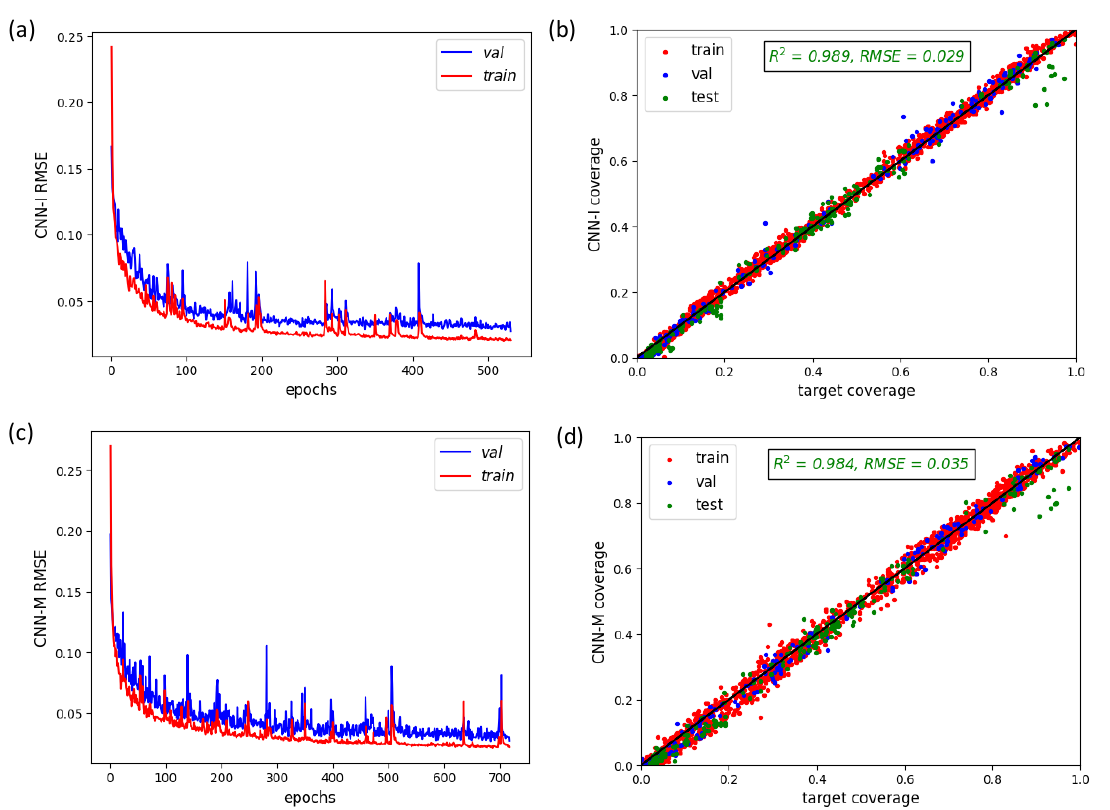}
         \caption{(a), and (c) is the root mean squared error value (RMSE) on the train and validation (val) data against the learning iteration (epochs) for the fine-tuned ResNet18 pretrained on the ImageNet data (CNN-I), the fine-tuned ResNet18 pretrained on the MicroNet data (CNN-M), respectively. (b) and (d) is the parity plots of the predicted and target coverage corresponding to (a) and (c), respectively. The $R^2$ and RMSE values in (b) and (d) are for the test set.}
         \label{fig:fine-tune}
\end{figure}

Finally, we examined the fine-tuning of the pretrained model to predict the crystal coverage.
This approach needs to be explored especially because we observe the significant impact data augmentation has on CNN model performance.
Fine-tuning is carried out for the ResNet18 pretrained on the ImageNet and another on the MicroNet.
These models are termed CNN-I and CNN-M, respectively.
As observed in the CNNsc, capturing the grokking effect is important in obtaining the optimally trained model; the training and validation losses were closely monitored, and the training halted once the minimal obtainable validation loss was reached.
The validation loss associated with the grokking point was determined by initial training of the models for a few thousand epochs.
The performance of CNN-I and CNN-M are quite similar, with CNN-I giving a marginally better result.
Both have accurate predictions on the validation and test set with $R^2$ value of 0.99 (see Fig.~\ref{fig:fine-tune} and Table~\ref{table:regression_results}). 

Interestingly, while a significantly better performance is observed from features extracted from the model pretrained on MicroNet than that from ImageNet, the fine-tuning shows the reverse.
This means that the filters pretrained on the MicroNet extract much more useful features from the AFM than those pretrained on the ImageNet.
However, the latter scenario seems to provide more generic image features in which case fine-tuning on sufficient target data has yielded a better result.
A nearly non-existent over-fitting, even on the held-out test data is noteworthy.
The excellent performance of CNN-I and CNN-M underscores the advantage of not just the transfer learning but also the data augmentations used with CNN to combat the over-fitting and producing models that have been accurately trained on our target data which share generic features learned from larger data sets used for the pretraining.

\subsubsection{Summary of Regression Results}

The results of all the regression models have been compiled in Table~\ref{table:regression_results}.
While comparable performance on training data can be obtained by all three learning paradigms, their test performance varies substantially.
Fine-tuning yielded the best results in this regard, followed by training from scratch, and then feature extraction.
However, this seems to have been largely a result of on-the-fly data augmentation, as our ablation study showed that removing this from the trained-from-scratch CNNsc led to a nearly triple test RMSE, making it the worst model.
Unfortunately, this approach could not be applied to the feature extraction strategy to improve its performance.
Between the two pretraining domains, there was no clear winner; ImageNet gave better performance in fine-tuning, while MicroNet was superior in feature extraction.
This is not an obvious result and may warrant further investigation regarding the nature of the pretrained filters.

    \begin{table}
        \centering
        \caption{RMSE and $R^2$ values for the predicted coverage on the train, validation (val), and test sets for models trained from scratch and through transfer learning.
        CNNsc and CNNsc* are the CNNs trained from scratch with and without on-the-fly data augmentation, respectively.
        MLP-I and MLP-M are the MLPs trained using the features extracted with ResNet18 architecture pretrained on ImageNet and MicroNet, respectively.
        CNN-I and CNN-M are the fine-tuning models of the ResNet18 architecture pretrained on ImageNet and MicroNet, respectively.
        The best performance in each row is shown in bold, including ties and near-ties.}
        \label{table:regression_results}%
        \begin{tabular}{l| c     c |      c       c  |     c       c}
       & \multicolumn{2}{c|}{\bf{From scratch}}       & \multicolumn{2}{c|}{\bf{Feature extraction}}&   \multicolumn{2}{c}{\bf{Fine Tuning}} \\\hline \hline
        RMSE& CNNsc& CNNsc*   & MLP-I & MLP-M &  CNN-I &  CNN-M\\
        \hline
        train & 0.018      & \textbf{0.013}  &  \textbf{0.012}  &  0.023 &  \textbf{0.013} & 0.022  \\
        val   & 0.039      & 0.120  &  0.098  &  0.047 &  \textbf{0.021}  &  0.030 \\
        test  & 0.041     & 0.121  &  0.101  & 0.054 &  \textbf{0.029}  &  0.035 \\ \hline \hline

                $R^2$& CNNsc& CNNsc*   & MLP-I & MLP-M &  CNN-I &  CNN-M\\
                \hline
        train & 0.997      & \textbf{0.998}  &  \textbf{0.998}  &  0.995 &  \textbf{0.998} & 0.995  \\
        val   & 0.984      & 0.855  &  0.904  &  0.978 &  \textbf{0.995}  &  0.991 \\
        test  & 0.979     & 0.818  &  0.873  & 0.963 &  \textbf{0.989}  &  0.984 \\ \hline
        \end{tabular}
    \end{table}

\subsection{Segmentation Models}

\begin{figure}
     \centering
     \includegraphics[width=\textwidth]{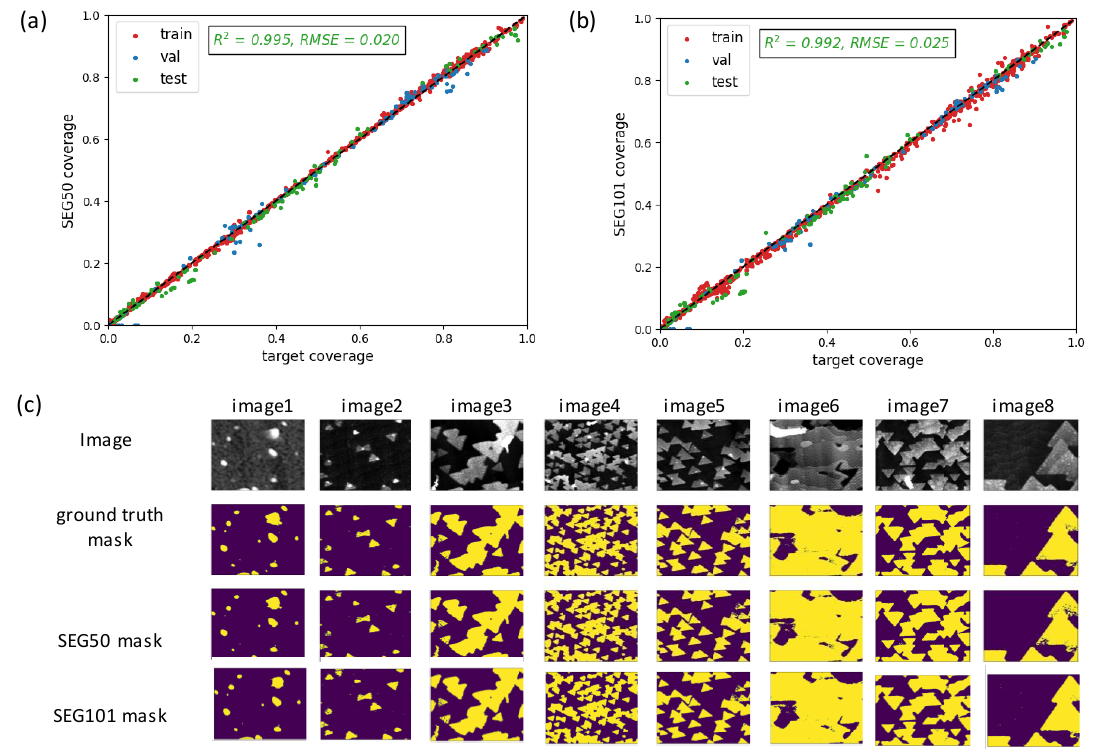}
         \caption{(a) and (b) are the parity plots of coverage predicted, using the segmentation model pretrained on the MicroNet, and the target coverage. The encoder for the SEG50 and SEG101 models are \texttt{SE\_ResNeXt-50\_32x4d} and \texttt{SE\_ResNeXt-101\_32x4d}, respectively. (c) are sample images, the corresponding ground truth mask, and the predicted mask by the SEG50 and SEG101 models. The $R^2$ and RMSE values are for the test set.}
         \label{fig:segmentation}
\end{figure}

 \begin{table}
        \centering
        \caption{The RMSE, $R^2$, and IOU values on the train, validation (val), and test data sets for the segmentation models. SEG18-I and SEG18-M use ResNet18 pretrained on ImageNet and MicroNet, respectively. \texttt{SE\_ResNeXt-50\_32x4d} (SEG50) and \texttt{SE\_ResNeXt-101\_32x4d} (SEG101) encoder are pretrained on MicroNet data.}
        \label{table:segmentation}%
        \begin{tabular}{l| c     c       c   |    c      c       c| c   c   c}
       & \multicolumn{3}{c|}{\bf{RMSE}}       & \multicolumn{3}{c|}{\bf{$R^2$}}  & \multicolumn{3}{c}{\bf{Average IoU (\%)}}\\\hline \hline
        & train& val   & test & train &  val &  test    & train &  val &  test\\
        \hline
    SEG18-I &0.017  & 0.021  & 0.022 & 0.997 & 0.995 & 0.994 & 89$\pm{21}$ &    88$\pm{26}$   & 90$\pm{13}$  \\
    SEG18-M &0.028  & 0.043  & 0.020 & 0.992 & 0.977 & 0.995 & 87$\pm{22}$ &    87$\pm{27}$   & 90$\pm{14}$  \\
        SEG50 & 0.007      & 0.024  &  0.020  &  0.999 &  0.993 & 0.995& 92$\pm{19}$ &    90$\pm{23}$   & 92$\pm{9}$  \\
        SEG101   & 0.013      & 0.020  &  0.025  &  0.998 &  0.997  &  0.992& 90$\pm{18}$ & 89$\pm{25}$ & 90$\pm{13}$ \\ \hline 
        \end{tabular}
    \end{table}
    
We now reframe the task as a binary segmentation, where the crystal (foreground) is separated from the substrate (background) and then counted to obtain the crystal coverage.
\texttt{SE\_ResNeXt-50\_32x4d} and \texttt{SE\_ResNeXt-101\_32x4d} encoders pretrained on MicroNet coupled with Unet++ decoders are termed SEG50 and SEG101, respectively.
While the ResNet18 encoder pretrained on the ImageNet and another on the MicroNet with both coupled with the Unet++ are termed SEG18-I, and SEG18-M, respectively.
As this is natively a segmentation problem, it is not surprising that these models can achieve excellent performance; all the segmentation models have a minimal improvement over the regression models as shown in Table~\ref{table:segmentation} and Fig.~\ref{fig:segmentation}. 
To be specific, the best model from the regression models, CNN-I (Fig.~\ref{fig:fine-tune} and Table~\ref{table:regression_results}) exhibits a test RMSE of 0.029, whereas SEG18-M and SEG50 both obtain 0.020 RMSE.

Based on the patches of the images, it seems that segmentation models provide higher performance in determining the crystal coverage than regression models.
Additionally, segmentation models offer the advantage of giving impressive performances even with a much smaller data set for training\cite{stuckner2022microstructure, seg1akers2021rapid, seg2azimi2018advanced} since each pixel is in effect a training data point.
In our present study, the total image patches used in the segmentation models are half of that used in the regression models.

In addition to the coverage value determination, segmentation models provide pixel-wise classification of the image, classifying each pixel in the AFM images of \ce{WSe2} samples as either belonging to the substrate or the crystal.
This has some additional utility in determining not only how much crystal is present, but its location in the micrograph.
The intersect over union (IoU) metric shows high performance even on the pixel-level classification task, with 92\% (SEG50) and 90\% (SEG101) IoU on held-out test images.
It is worth noting that similar performances are observed on both the train and test sets, indicating low memorization.
This level of generalization, despite the held-out test set samples being grown at different conditions and/or obtained at different imaging conditions, underscores the potential of the models to produce reliable results in practical applications.

\subsection{Inference on Full Images}

\begin{figure}
     \centering
     \includegraphics[width=\textwidth]{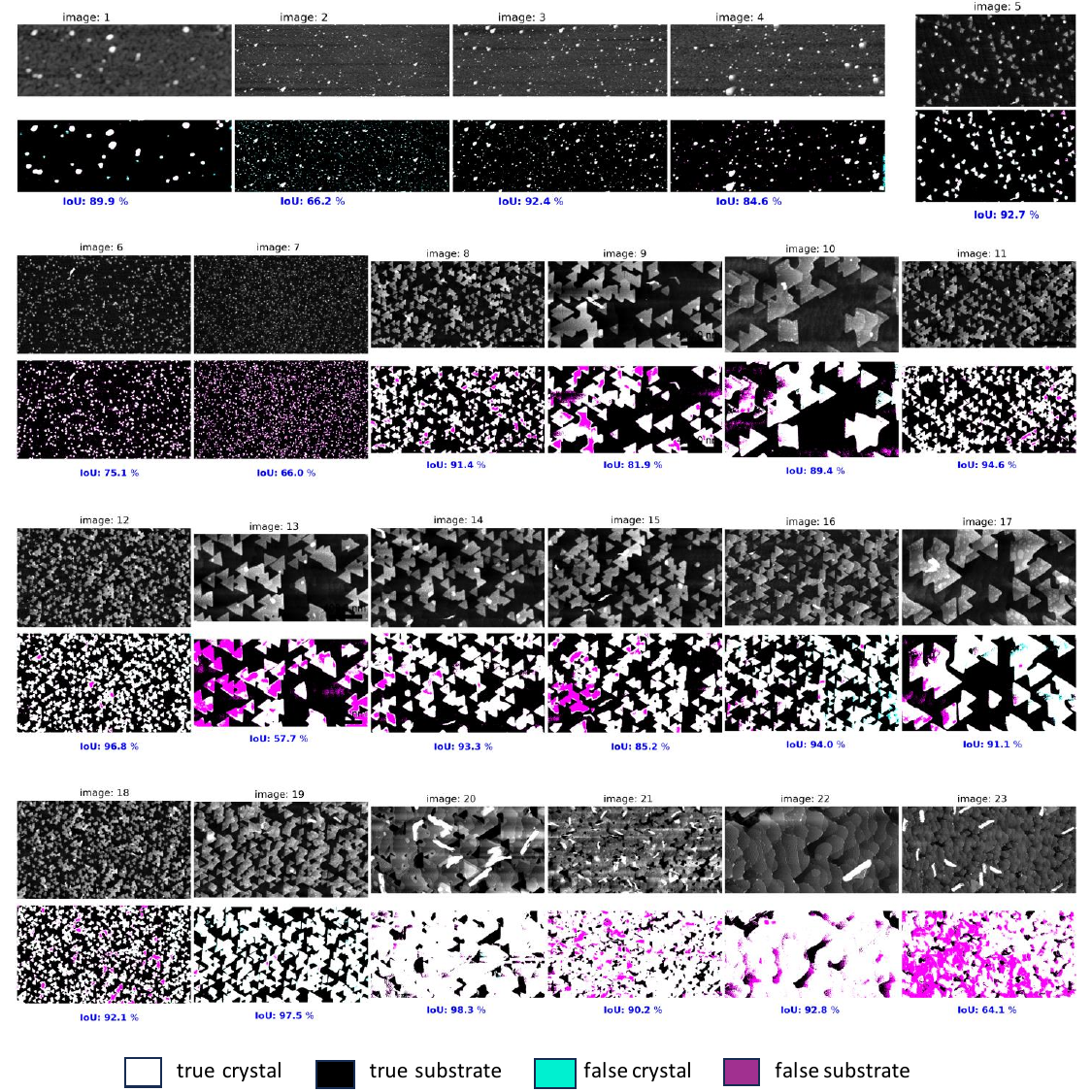}
         \caption{The original (whole) images in the hold-out test set (first rows), and the pixel-wise classification, as either belonging to crystal or substrate (second rows) obtained from the SEG50 model. The intersection over union (IoU) accuracy for each image is given below the classification.}
         \label{fig:inference}
\end{figure}

\begin{figure}
     \centering
     \includegraphics[width=\textwidth]{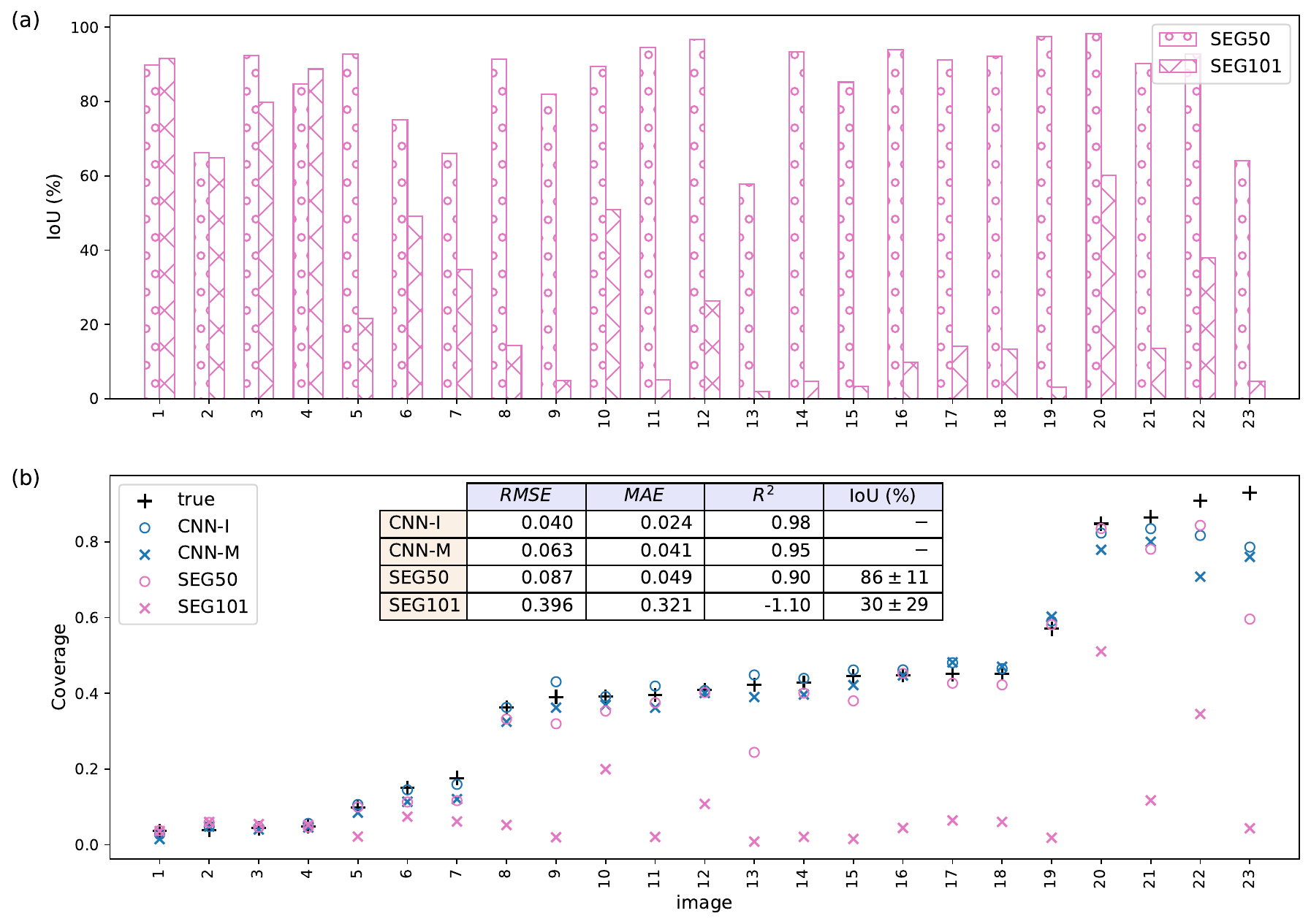}
         \caption{Coverage analysis and segmentation of the original (whole) test images. Results obtained using the segmentation models, SEG50 and SEG101, and the best regression models, CNN-I and CNN-M are shown. The S/No. corresponds to the image \# shown in Fig.~\ref{fig:inference}.}
         \label{fig:inference2}
\end{figure}

The test set discussed in the previous sections is based on patches created from the full image test set. 
However, it is important to characterize the held-out test set in its original full image format, as this is the real measure of the practical value of our trained models.
For this test, we are using SEG50 and SEG101 and only the best regression models: CNN-I and CNN-M.
While SEG50 gives the best performance on the held-out test set among the segmentation models, SEG101 and SEG18-I give similar results (Table \ref{table:segmentation}).

The full images were padded such that they match the exact multiple of model training patch size, $224 \times 224$ and $512 \times 512$, for regression and segmentation, respectively, or the last row/column is lost.
The tiles (with the same sizes as those used in training the models) are then obtained from the full images and the coverage and segmentation are predicted using the trained models.
For CNN-I and CNN-M, the predicted coverage for each tile is multiplied by the size of the tile to obtain the number of pixels with the value above the threshold for the crystal.
The pixel values above the threshold are added for all the tiles from the same full image.
The crystal coverage of a given full image is then obtained by dividing the sum of the number of pixels above the crystal threshold from all the tiles by the size of the full image (the total number of pixels in the full image).
Meanwhile, for the SEG50 and SEG101, the resulting segmented tiles are concatenated and the artificial padding added is removed.
The coverage label is then obtained based on the concatenated segmentation mask. 

For the 23 held-out test images which were grown with different growth parameters and/or obtained at different imaging conditions than the train and validation sets (Table~\ref{fig:data}), the performance of the models is not as good as on the patched images for any model.
The regression models are at least 30\% worse while the segmentation models are at least four times worse -- this means that the regression models outperform the segmentation models in practice despite worse test performance on image patches (Figures \ref{fig:inference} and \ref{fig:inference2}).
The results obtained from SEG50 are mostly consistent with the results on image patches with an average IoU accuracy of 86\% compared to 92\%.
Except for a few cases such as the image \#6, 15, and 19, less than 10\% errors are typical for both the coverage and the IoU.

In contrast, the SEG101 performed quite poorly, despite being a similar architecture compared to the SEG50, which is surprising because both models give comparable performance on the patched images.
The fact that SEG101 gave the best result on the first 4 images, which are the same size but different from the rest of the test set, provides a clue as to why the model performs poorly on most of the images as well as the SEG50's lower accuracy on the full images compared to the patches.
Creating the tiles for the full image inference requires processing that could result in the loss of some parts of the original images.
The resizing involved in the patches created for training the models is also inevitably not the same as that for the tiles.
The sensitivity of the different models to the different image processing and the image morphological features have therefore resulted in the observed variation in the model performances.
Also worthy of note is the fact that significant variations in the segmentation model performances have been observed depending on the encoder and/or decoder architecture.\cite{stuckner2022microstructure} 

Overall, the results on full images show an important distinction between the training protocol and the real-world application of CNNs.
Deep CNNs such as SEG101 may not be robust in practical micrograph analysis despite excellent performance even on held-out test data due to the image augmentation scheme.
Meanwhile, even though the calculation of crystal coverage is natively a segmentation problem, the regression models perform well on the full images, suggesting that they may be more robust to changes of scale, dimension, or other factors compared to the segmentation models.

\section{Conclusion}

In this study, we conduct a comprehensive analysis of crystal coverage (the proportion of the substrate covered with grown crystal) in \ce{WSe2} thin film atomic force microscopy (AFM) micrographs using regression and segmentation models.
Regression models were trained to predict the monolayer crystal coverage from image patches.
Models were trained from scratch and using transfer learning from ResNet pretrained on ImageNet and MicroNet.
MicroNet consists of grayscale micrographs while ImageNet is made up of the macroscale color images of natural objects.
For transfer learning, both feature extraction and fine-tuning approaches were used. 

Our analysis revealed that the CNN models trained from scratch outperform MLP models trained on features extracted from the pretrained models, while fine-tuning gave the best performance with up to 0.99 $R^2$ value on the held-out test set. 
Interestingly, while a significantly better performance is observed from feature extraction using MicroNet than that from ImageNet, the fine-tuning shows the reverse.
This means that the filters pretrained on the MicroNet extract more useful features from the AFM than that pretrained on the ImageNet.
However, the latter scenario seems to provide more generic image features in which case fine-tuning on sufficient target data has yielded a better result.

Beyond the prediction of crystal coverage over entire patches, segmentation models provide pixel-wise classification of the image, classifying each pixel in the AFM images of \ce{WSe2} samples as either belonging to the substrate or the crystal.
This has some additional utility in determining not only how much crystal is present, but its location in the micrograph.
Based on the patches of the images, the segmentation models provide higher performance in determining the crystal coverage than regression models.
The intersection over union (IoU) metric shows high performance even on the pixel-level classification task, with up to 92\%  IoU on held-out test images.

The results on full images show an important distinction between the training protocol and the real-world application of the models.
Contrary to the results from image patches, the regression models performed better than the segmentation models at predicting the monolayer crystal coverage of the full images of the held-out test set, giving the $R^2$ values of 0.98 and 0.90, respectively, from the best models.
The average IoU on the full held-out test images reduced to 86\% from the 92\% obtained for the patch images.
Our finding suggests that the regression models may be more robust to changes in scale, dimension, or other factors compared to the segmentation models. 
Overall, these results highlight the efficacy of machine learning for automated, high-throughput sample characterization, demonstrating its potential for accelerating the high-throughput development of chalcogenides for technological applications.
At the same time, it provides practical guidelines for implementing standard computer vision workflows in real-world materials characterization applications.

\section*{Acknowledgments}
This study is based upon research conducted at The Pennsylvania State University Two-Dimensional Crystal Consortium – Materials Innovation Platform (2DCC-MIP) which is supported by NSF cooperative agreement DMR-2039351.

\section*{Data Availability}
The raw data required to reproduce these findings are available to download from Ref.\cite{WSe2data}
The processed data required to reproduce these findings are available to download from Ref.\cite{moses_2024_10784189}
Codes used to generate the results reported are available at https://github.com/reinhart-group/wse2$\_$coverage.


\bibliography{main} 

\end{document}